\documentclass[twocolumn]{pasj00}
\usepackage{amsmath} 
\usepackage{graphicx}
\SetRunningHead{S. Onodera et al.}{Gas Dynamics in NGC 4501} 
\title{Virgo High-Resolution CO Survey\\ IV. Spiral-Driven Gas Dynamics in 
the Non-Barred Seyfert Galaxy NGC 4501}  
\author{Sachiko \textsc{Onodera},$^1$ Jin \textsc{Koda},$^{1,2,3}$ Yoshiaki 
\textsc{Sofue},$^1$ and Kotaro \textsc{Kohno}$^1$}  
\affil{$^1$Institute of Astronomy, The University of Tokyo, Mitaka, Tokyo 
181--0015} 
\affil{$^2$ Nobeyama Radio Observatory, National Astronomical 
Observatory, Minamimaki, Minamisaku, Nagano 384--1305} 
\affil{$^3$ ALMA Project Office, National Astronomical 
Observatry, Mitaka, Tokyo 181--8588}
\email{sonodera@ioa.s.u-tokyo.ac.jp} 
\Received{2003 October 28}
\Accepted{2004 April 21} 
\Published{$\langle$publication date$\rangle$} 
 
\begin{document} 
\maketitle 
 
\begin{abstract} 
We report on high-resolution interferometer observations 
 of the $^{12}$CO({\it J}=1--0) emission in the  
 central 5 kpc region of the Seyfert 2 galaxy NGC 4501. The 
 observations were made using the Nobeyama Millimeter Array  
 during a long-term CO line survey of Virgo spirals. 
The major features are: (1) a nuclear concentration with a
 radius of  
 $r\sim 5''$ (390 pc), which is resolved into double peaks, and 
 (2) spiral arms which extend out from the nuclear region. 
 The nuclear component has a mass of $1.3\times 10^8\: M_\odot$, which 
 corresponds to $\sim$3.5\% of the dynamical mass, 
 and shows a slight non-circular motion. The double peaks are separated
 by $\sim \timeform{4''.7}$   
($370\: \rm pc$), and located on the root of
 optical spiral arms in a HST image. 
 The gas arms are 
 associated with the spiral dust lanes, and are linked to the central 
 double peaks. The 
 non-circular motions along the molecular arms indicate the fact 
 that the gas is driven by the  
 density wave, rather than the stochastic processes. 
 We calculated the gas cloud 
 orbits in a stellar spiral potential, and explained 
 the observed CO spiral arms and non-circular motions.
 We suggest that the central gas condensation arises from  
spiral-driven gas transfer.   
We estimated and compared the effect of two possible mechanisms of 
angular-momentum transfer: galactic shock, and gravitational torques.
We discuss that the galactic shock is dominant.  
\end{abstract} 
 
\section{Introduction} 
Gas dynamics in disk galaxies is closely related to the stellar 
 potential. 
Gas dynamics in barred galaxies has been well 
studied in particular, and many observed features have been successfully explained. 
Bars are thought to be the most viable candidates for
 fueling the central activity, transferring mass from large to small 
 scales (e.g. \cite{simkin80}; \cite{shlosman89}; \cite{athana92}). 
Recent statistics report circumstantial evidence of radial gas 
inflow as a result of bar-driven gas dynamics.  
Metallicity gradients in barred spirals are systematically flatter 
than that of unbarred ones, and they become shallower when the bar length 
and the bar ellipticity increase \citep{martin94}.  
The degree of gas condensation to the central kiloparsec  
is higher in barred galaxies than in unbarred ones \citep{Sakamoto99b}. 

However, the central gas condensation is also prevalent in 
non-barred galaxies \citep{Sakamoto99a}, and some studies claim 
that the level of the central activity does not correlate with the presence of 
large-scale bars (\cite{Ho97b}; \cite{MR97}). 
Their results suggest that gas inflow may occur both in barred and 
non-barred spiral galaxies. Although a theoretical study of  
the gas dynamics in spiral galaxies with tightly wound arms was in full flourish 
several decades ago (e.g. \cite{Fujimoto68}; \cite{Roberts69}),  
its application for observational data and discussions within the context 
of radial mass transfer has not been thoroughly pursued
compared to that in barred galaxies.   
 
We observed a non-barred CO-luminous galaxy, NGC 4501,  
during the Virgo high-resolution CO survey (ViCS: 
\cite{sofue03a}), using the Nobeyama Millimeter Array with the 
highest resolution ever achieved for this galaxy.  The prominent  
molecular arm structure, non-circular motion, and centrally-condensed 
double peaks motivated our work on the gas dynamics 
in the spiral potential and possible inflow mechanism to the center. 
 
NGC 4501 is one of some non-barred galaxies 
that have a relatively high degree of central gas condensation, similar to 
those of barred galaxies \citep{Sakamoto99b}. 
It is an SAb galaxy 
(\cite{RC3}, hereafter RC3) at a distance of $16.1\rm\, Mpc$  
(\cite{Ferrarese96}; $1''$ corresponds to $78\: \rm pc$) 
located about \timeform{2D.0} (0.56 Mpc) northeast of  
the center of the Virgo cluster, M 87. It hosts a weak 
Seyfert 2 nucleus \citep{Ho97a}. 
It is slightly H\emissiontype{I} 
 deficient \citep{Chamaraux80}, and has a high CO luminosity 
 and a high IRAS flux density  
\citep{Stark86}.  
 While it is not classified as a grand-design spiral galaxy in 
{\it B}-band image (arm class 9; \cite{EE87}), it 
shows continuous two-armed spiral in the {\it K$'$}-band
within $r\lesssim 3\: \rm kpc$ (\cite{Elmegreen99}, see their figure 1).  
Moreover, \citet{carollo98} noted the 
spiral-like dust lanes 
down to the nucleus in the F606W (wide $V$-band) image of HST WFPC2. 
The basic properties of NGC 4501 are listed in table 1. 

 \begin{table}[t] 
 \begin{center} 
 \caption{Parameters of NGC 4501.}\vskip 2mm 
 \label{tab:t1} 
 \begin{tabular}{lcc} 
 \hline 
 \hline 
 Parameter & Value & Source$^*$\\ 
 \hline 
 Hubble type & SA(rs)b & 1\\ 
 Nuclear activity & Type 2 Seyfert & 2\\ 
 Adopted distance (Mpc)  & 16.1 & 3\\ 
 P.A.(isophotal)($\degree$) & 140 & 4\\ 
 Inclination($\degree$) &58 & 4\\  
 $D_{25}\times d_{25}$ & $\timeform{6'.9}\times \timeform{3'.7}$ & 1\\ 
 $V\rm{_{sys}\; (\: \rm km\,s^{-1})}$ & 2263 & 4\\ 
 $B_{\rm total}^0$ (mag) & 9.86 & 1 \\ 
 $S_{\rm CO}(45'')$(K$\: \rm km\,s^{-1}$) & 12.0$\pm$4 & 4\\ 
 Linear scale ($\rm pc\, arcsec^{-1}$) & 78.1 & \\ 
 \hline 
 \end{tabular} 
\end{center} 
 \footnotesize{$^*$ (1) \citet{RC3} (RC3); (2) Ho 
 et al. (1997a); (3) 
  Ferrarese et al. (1996); (4) Kenney and Young (1988)}\\ 
\end{table}  
 
\normalsize 
 
In this paper, we show that the gas dynamics in the central 
region of this galaxy is governed by a stellar spiral 
potential, 
 and suggest the possibility that the observed central 
features can be the outcome of radial inflow due to the spiral arms. 
High-resolution $^{12}\rm CO$({\it J}=1--0) 
observations of NGC 4501 are presented in section 2. 
The results from the observations are presented in section 3. The CO 
distribution and the velocity field are modeled based on observations in section 
4. Possible 
mechanisms of spiral-induced radial gas inflow are discussed in section 5. 
We summarize our conclusions in section 6. 
A detailed description of the damped-orbit 
model in the spiral potential is shown in appendix 1. An estimation
 of the change in the angular momentum  
with shocks and torques is presented in appendix 2. 
\section{Observations and Reduction} 
\subsection{CO(J=1--0) Observations with NMA} 
We carried out observations of NGC 4501 in the $\rm ^{12}$CO(\it 
 J=\rm 1--0) line using the Nobeyama Millimeter 
 Array (NMA) at the Nobeyama Radio Observatory (NRO) during the course of 
 a long-term project, the Virgo high-resolution CO Survey (ViCS; Sofue et 
 al 2003). The  
 observations were made from 2000 December 
 to 2002 March for a single pointing center at ($\alpha_{\rm J2000}$, 
$\delta_{\rm J2000}$)=(\timeform{12h31m59s .14}, \timeform{+14D25'12''.9}). The NMA 
 consisted of six  antennas, each had a diameter of 10 m, providing a 
 FWHP of about 
 $65''$ at 115 GHz. We used  
 the spectro-correlator Ultra-Wide-Band Correlator (UWBC: \cite{Okumura00}), in a mode of 256 channels covering 512 MHz. One channel (2 MHz) 
 corresponded to $5.2\: \rm km\,s^{-1}$ at the observing frequency. 
 Quasar 3C 273 was observed every 20 minutes to correct for any 
 instrumental gain variations and the bandpass 
 response. Continuum observations of Mars were used to calibrate the 
 flux scale. We calibrated the absolute flux scale in each sideband in each 
 observation: once in 2000, three times in 2001, and once in 2002. We 
 compared the noises in two sidebands and used the lower sideband (102 
 GHz) data in all 
 flux calibrations. The flux density of 3C 273 was roughly constant at 9.6 Jy,  
 within $\pm\sim 20$\%. No significant flux variation was observed 
 in all of our runs. 
The chopper-wheel method was used to correct for any 
 atmospheric transmission losses and elevation-dependent gain variations. 
 We used AB, C, and D array configurations of the NMA to achieve a 
 resolution of less than $2''$. The visibility data 
 cover projected baselines from 10 to 341 m. Partially shadowed 
 observation data were deleted, and the minimum baseline length 
 projected onto the sky was the diameter of 
 each antenna. 
 This resulted in the central 
 hole in the {\it u--v} plane, which restricts the largest detectable size of 
 objects to about $54''$. The observation parameters are listed in table 2. 
 
\begin{table}[t] 
\begin{center} 
\caption{Observation Parameters}\vskip 2mm 
\begin{tabular}{lc} 
\hline
Observed center frequency (GHz) & 114.407 \\ 
Array configurations & AB, C, and D \\ 
Observing field center &\\
 \citep{Sakamoto99a}: & \\ 
 \hskip 10mm $\alpha$(J2000) & \timeform{12h31m 59s.14}\\ 
 \hskip 10mm $\delta$(J2000) & \timeform{+14D25'12''.9}\\  
Frequency channels & 256 \\ 
Total bandwidth (MHz) & 512 \\ 
Velocity coverage ($\: \rm km\,s^{-1}$) & 1342 \\ 
Velocity resolution ($\: \rm km\,s^{-1}$) & 5.24 \\ 
Amplitude and phase calibrator & 3C 273 \\ 
Primary beam (\timeform{''}) & 65 \\ 
\hline 
\end{tabular} 
\end{center} 
\end{table} 
\subsection{Reduction} 
 The raw visibility data were calibrated with the NRO/UVPROC2 software 
 \citep{tsutsumi97} and 
 mapped with the NRAO/AIPS package. 
We applied the CLEAN procedure with two ways of weighting (natural and 
 pure uniform), 
and obtained three-dimensional data cubes (R.A., Decl, 
 $V\rm_{LSR}$) in low ($5''$) and high ($2''$) spatial resolutions. The 
 synthesized beam  
 parameters, velocity resolutions and rms noises for these 
 maps are listed in table 3. The fraction of the total flux recovered by 
 our NMA data cube was estimated by comparing with 
 single-dish observations using the FCRAO 14 m telescopes \citep{KY88}. 
For this comparison, the data cube was corrected for the primary beam 
 attenuation of the NMA antennas, convolved with the single-dish beam 
 (FCRAO, $45''$), and sampled at the pointing centers of the single-dish 
 observations to obtain a value comparable to the single-dish flux.  
 The NMA cube was found to recover about 78\% of the single-dish flux 
 in the central $45''$. Note, however, the primary beam correction was 
 not applied to the maps shown in this paper.  

\begin{table*}[t] 
\begin{center} 
\caption{Parameters of Maps.$^*$} 
\begin{tabular}{cccccccc} 
\hline\hline 
 & & \multicolumn{2}{c}{Beam} & & Velocity  
 & rms noise & $T\rm _b$ for\\ 
 \cline{3-4}  
Resolution & Weighting & FWHM & P.A.&& Resolution  
&$\sigma$& $\rm 1\, Jy\, beam^{-1}$ \\ 
 & &(\timeform{''}) & (\timeform{D}) &&($\: \rm km\,s^{-1}$) 
 &($\rm mJy\,beam^{-1}$)&(K) \\ 
\hline 
Low & Natural & $5.6 \times 3.7$  & 160 && 10.5 & 14 & $\>$ 4.5 \\ 
High & Pure uniform & $1.8\times 1.7$  & 138 && 41.9 & 16 & 30.5\\ 
\hline
\multicolumn{8}{l}
{ \footnotesize \footnotemark[$*$] The map centers are set at derived dynamical center:}
\\
\multicolumn{8}{l}
{ \footnotesize ($\alpha_{{\rm J}2000}$, $\delta_{{\rm J}2000}$)=(\timeform{12h31m59s.12}, \timeform{+14D25'13''.3}) and  
derived systemic velocity 2261 $\: \rm km\,s^{-1}$.}
\end{tabular} 
\end{center} 
\end{table*} 

Figures \ref{fig:low} and \ref{fig:high} display low- and 
 high-resolution integrated  
 intensity maps and velocity  
 fields in the central  
 region of NGC 4501 ($90''\sim 7.0$ kpc square for figure \ref{fig:low}, 
 $14''\sim 1.1 \: \rm kpc$ for figure \ref{fig:high}). The crosses in figure 
 \ref{fig:low} and figure \ref{fig:high} 
 indicate the major and minor axes of NGC 4501 (P.A.=$140\degree$).  
The ellipse in figure \ref{fig:low} indicate a projected circle of 
 diameter $75''$  
 (5.9 kpc), inclined  
 with the position angle and inclination ($58\degree$) 
 of this galaxy. The size of the crosses in high-resolution maps is 
 $5''$ ($390\: \rm pc$). The center of all crosses and ellipses in figures 
 \ref{fig:low} and \ref{fig:high} are set at the 
 dynamical center derived from the velocity field (see below).  
 
 Figure \ref{fig:ch} shows channel maps with an interval of 4 MHz 
 ($10.4\: \rm km\,s^{-1}$) in the same  
 region. 
 Significant emission ($>3\sigma$, $1\sigma =14\rm\, mJy\, beam^{-1}$) 
 was detected in 46  
 adjacent channels within the velocity range of ${\it V}_{LSR}=2044-2513 \, {\rm km\, 
 s^{-1}}$ ($\Delta V=269\, \rm km\, s^{-1}$). In order to check the 
 contribution of the continuum flux, we examined the lower sideband (102 GHz) 
 data that can be regarded as being emission-free. We summed up our data
 over the full bandwidth (512 MHz); no significant continuum flux ($>3\sigma$, $1\sigma =1.3 \rm\, mJy\, beam^{-1}$) was detected.  

We obtained kinematical parameters from the low-resolution ($5''$) 
velocity field using the  
AIPS/GAL package. The dynamical center (table 2) was determined using 
the Brandt rotation curve model \citep{Brandt65}. For fitting, we used the central 
 $r<5''$ region where the velocity field can be regarded as being nearly 
 symmetric against the dynamical center. 
The initial guess 
 of the parameters affected the result within $r<1''$. 
 The centers of CO maps 
 and PV diagram are set at this derived dynamical center: ($\alpha_{\rm J2000}, 
 \delta_{\rm J2000}$)=(\timeform{12h31m59s.12},\timeform{+14D25'13''.3}) 
 and the systemic velocity 
 $V_{\rm sys}=2261 \: \rm km\,s^{-1}$.  
A GAL fit also gave the inclination and position angle of the 
 inner disk. However, these quantities are dependent on the non-circular 
 rotation, which are not negligible in the central regions. Therefore, 
 we adopted the inclination ($ 
58\degree$) and position angle ($140\degree$) determined from optical 
 images of the entire stellar disk. 
 
 Figure \ref{fig:rad} shows the radial profile of the CO-line 
 intensities in units of 
 $\rm Jy\: \rm km\,s^{-1}\,arcsec^{-2}$, which is corrected for the
 inclination. At first, low-resolution integrated-intensity map 
 was corrected for 
 primary-beam attenuation. We then applied the AIPS task IRING to 
 azimuthally average the integrated intensities in each annulus with 
 corrections for the inclination. We applied the 
 central positions derived above and the inclination and position angle 
 provided from optical observations (table 1). 
Figure \ref{fig:pv} displays a position--velocity diagram (PV diagram) along the 
 optically defined major axis (P.A.=$140\degree$) with a slit width of
 $5''$ and $60''$, respectively. The velocities have not been corrected for 
the inclination. 
\subsection{Supplementary Data} 
We obtained optical photographs of NGC 4501 from archives for a
comparison. We overlaid low-resolution CO map contours on a 
Digitized Sky Survey $B$-band image in order to see the 
global pattern (figure \ref{fig:B}). 
 We also overlaid high-resolution CO map contours on an unsharp-masked 
 image made from the HST WFPC2 observations at the F606W (wide 
 \it V\rm -band) filter \citep{carollo98}, in order to see the 
 innermost $r<500\: \rm pc$ region with a higher resolution (figure \ref{fig:HST}).    
\section{Results} 
The obtained CO map (figure \ref{fig:low}) shows spiral arms 
and a 
strong central condensation. 
the central CO condensation is resolved into double 
peaks in the higher resolution map (figure \ref{fig:high}). We describe these 
components in the following subsections.  
\subsection{Main Disk} 
\subsubsection{Distribution in the main disk} 
The entire CO disk of NGC 4501 has an exponential profile of scale 
 length \timeform{0'.7}, 
 and it continuously extends to $r\sim \timeform{3'.8}$ \citep{KY88}. 
 We now discuss the central $r\lesssim 40''(3.1\: \rm kpc)$ region of the CO 
 main disk (figure \ref{fig:low}, {\it Left}). 
 The radial profile is nearly exponential (figure \ref{fig:rad}), and  
 the emission enhancements due to  
the molecular spiral arms are distinct (subsection 3.2). 
\subsubsection{Kinematics of the main disk} 
\begin{figure*}[t]
\begin{center}
\FigureFile(160mm,100mm){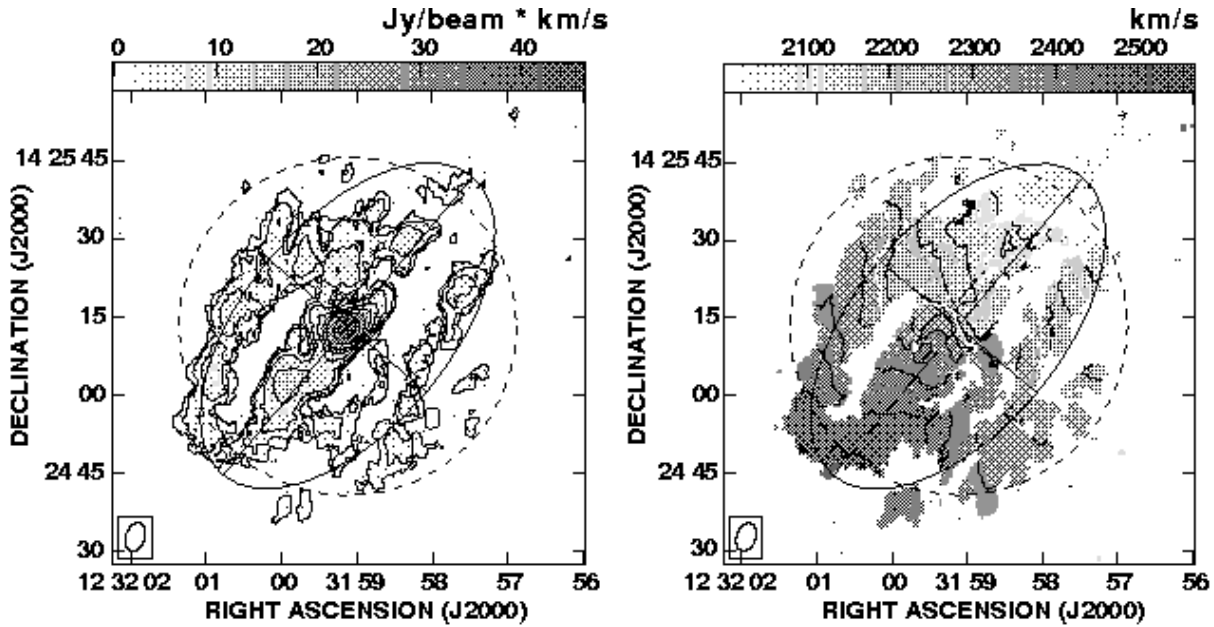} 
\end{center} 
\caption{Low-resolution ($\timeform{5''.6}\times \timeform{3''.7}$)  
CO(1--0) intensity map and 
 velocity field of NGC 4501. The  
 synthesized beam is shown in the lower left-hand corner. The 
 primary-beam attenuation is not corrected. The clip level was 
 $4\sigma$ ($1\sigma=14\rm \, mJy\, beam^{-1}$ in each channel) to make these 
moment maps. The circle represents the FWHP of NMA ($65''$). 
 The crosses indicate the orientations of the major and minor axes of NGC 
4501   
 (P.A.=$140\degree$). The center of maps, crosses and ellipses are set at 
 the dynamical center (table 2). The ellipse indicates a projected 
 circle of a diameter of $75'' (5.9\: \rm kpc)$  
inclined with the position angle ($140\degree$) and the inclination  
($58\degree$).  
Left: Intensity map in the central 
 $90''(\sim 7.0\: \rm kpc)$ squared region. The contours are 
 drawn at 5, 10, 20, 30, 40, 60, 80\% of the peak value 
 $46\: \rm Jy\, beam^{-1}\, \: \rm km\,s^{-1}$. The rms noise is $1\sigma=1.0\:\rm Jy\rm \,  
beam^{-1}\: \rm km\,s^{-1}$.  
 $1\:\rm Jy \, \rm beam^{-1}$ corresponds to 4.51K. 
 Right: 
 Velocity field. The contours  
 are drawn every 50$\: \rm km\,s^{-1}$ in the range of 2050--2550$\: \rm km\,s^{-1}$. The white line 
 indicates the systemic velocity, 2261$\: \rm km\,s^{-1}$.}  
\label{fig:low} 
\end{figure*}

The velocity field (figure \ref{fig:low}, Right) shows 
rotation of the main disk with the north--west side approaching and the
south-east side receding. It represents non-circular motions 
superposed on 
regular rotation of the disk. The non-circular motions appear as a
distortion of the isovelocity contours. The isovelocity contours around 
the spiral arms of $r\sim 2.9\: \rm kpc$ (the ellipse in figure \ref{fig:low}) 
indicate that 
the near (northeastern) sides of the arms are receding, and the far 
(southwestern) sides are 
approaching, as compared with pure circular motions.  
The non-circular motions also appear in $r\lesssim 10''$ 
as a Z-shaped pattern of the isovelocity contours. 
They can be attributed to spiral arms (subsection 3.2). The channel maps show 
the rotation of the disk and non-circular motion along the spiral arms as 
well (figure \ref{fig:ch}). 
The map centers were set at the derived dynamical center (section 2). 
Figure \ref{fig:rc} shows the 

\onecolumn

\begin{figure*}[b] 
\begin{center} 
\FigureFile(160mm,100mm){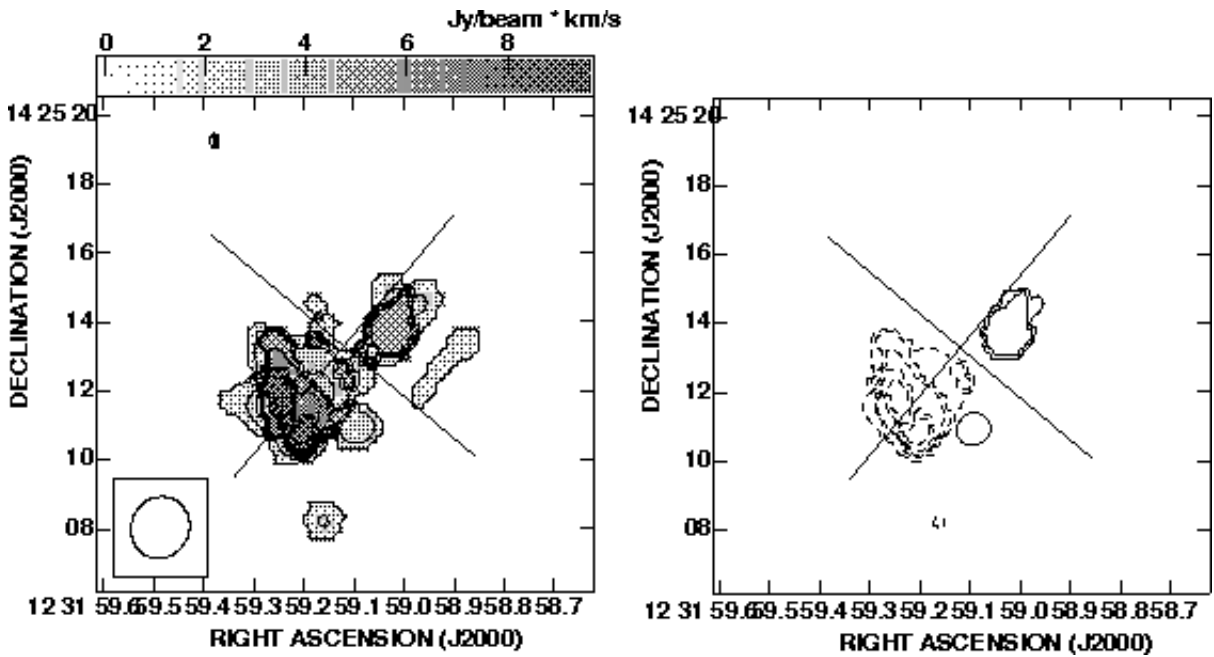} 
\end{center} 
\caption{High-resolution 
 (\timeform{1''.8}) CO(1--0) intensity maps and velocity field of NGC 4501. The 
 synthesized beam is shown in the lower left-hand corner. The 
 primary-beam attenuation is not corrected. The clip level was 
 $4\sigma$ ($1\sigma=16\rm\, mJy\, beam^{-1} $) to make these 
 maps. The crosses indicate the dynamical center and the orientation of 
the galactic major and minor axes. Their size are $5''(390\: \rm pc)$. Left:  
 Intensity map in the central $14''(\sim 1.1\: \rm kpc)$. 
 The contours are drawn at 30, 40, 50, 60, 70, 80, 90\% of the peak 
 value $9.6\: \rm Jy\,beam^{-1}\: \rm km\,s^{-1}$. The rms noise is  
$1\sigma=2.1\:\rm Jy beam^{-1}\: \rm km\,s^{-1}$. $1\:\rm Jy \, \rm beam^{-1}$ corresponds to 30.5K. 
Right: Intensity map representing velocity field,  
in which approaching 
 ($V<V_{\rm sys}$: solid lines) and receding ($V>V_{\rm sys}$: broken lines) 
 components are distinguished. The contours are drawn at 30, 45, 60, 75, 
 90\% of the peak value.}  
\label{fig:high} 
\end{figure*} 
\twocolumn

\onecolumn
\begin{figure*} 
\begin{center} 
\FigureFile(160mm,320mm){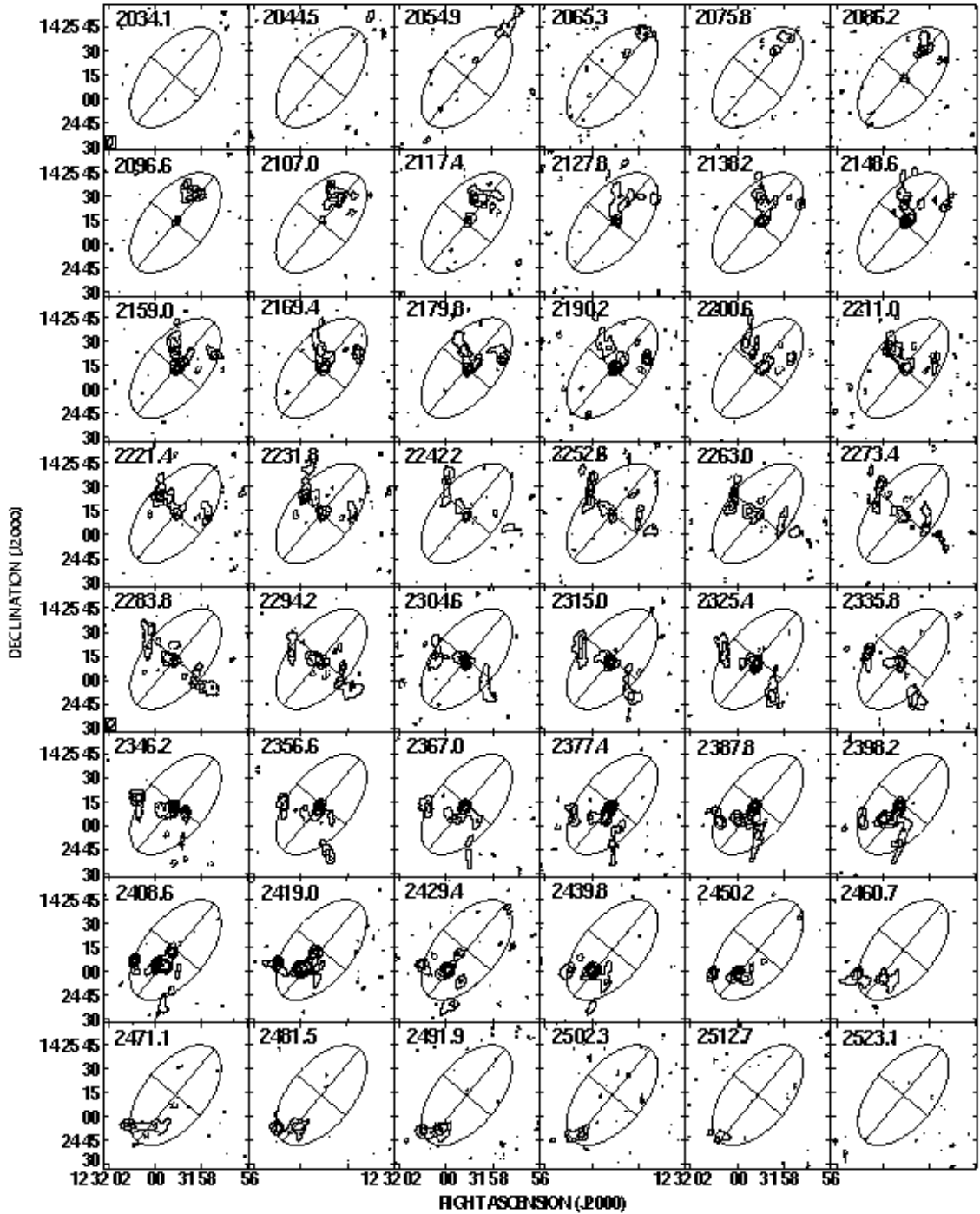} 
\leavevmode 
\end{center} 
\caption{Channel maps in the CO(1--0) emission. The ellipse is the same as 
 that of figure 1. The channels have an interval of $10.4 \: \rm km\,s^{-1}$. Their 
 central velocities ($V_{\rm LSR}$ in $\: \rm km\,s^{-1}$) are labeled at the upper 
 left corners. The contour levels are -3, 3, 6, 9, 12, 15 $\sigma$, where 
 $1\sigma =14\rm \,mJy\,beam^{-1}$. The negative contours are dotted. No  
primary-beam 
 correction was applied. The velocity distortion along the spiral 
 arms are apparent.} 
 \label{fig:ch}	 
\end{figure*} 
\twocolumn
$\!\!\!\!\!\!\!$inclination-corrected rotation 
 curve. 
 The rotation curve was derived from the position--velocity diagram along 
the major axis with a slit width of $3''$ using an 
 iteration method  
developed by \citet{TS02}. This method determines a rotation 
curve so that it can reproduce the observed position--velocity 
 diagram.  

The rotation velocity shows a steep rise in the central $5''$ to $v\sim 
200\: \rm km\,s^{-1}$, and then gradually rises to $\sim 290\: \rm km\,s^{-1}$ at $r\sim 35''$. This 
velocity is similar to the rotation velocity of the outer flat part 
\citep{nishiyama01}, but slightly excesses the H\emissiontype{I} rotation 
velocity,  
$268\: \rm km\,s^{-1}$, at 20\% of the peak flux \citep{cayatte90}.   
\subsubsection{Mass and surface brightness of the molecular gas} 
The mass of the molecular gas $M_{\rm H_2}$ is estimated from 
the total CO-line flux, $S_{\rm CO}$, 
adopting a Galactic $N_{\rm H_2}/I_{\rm CO}$ conversion factor, 
$X_{\rm CO}$, and the galactic distance, $D$, as 
\begin{eqnarray} 
\left(\frac{M_{\rm H_2}}{\: M_\odot}\right)=7.2\times &&10^3\left(\frac{D}{\rm 
 Mpc}\right)^2\left(\frac{S_{\rm CO}}{\rm 
 Jy\,km\,s^{-1}}\right)\nonumber \\
&&\times\left[\frac{X_{\rm CO}}{1.8\times  
 10^{20}\rm cm^{-2}\,(K\,km\,s^{-1})^{-1}}\right].  
\end{eqnarray} 
The gas surface densities are independent of the adopted distance to the 
galaxy. If we assume the hydrogen mass fraction 0.707 \citep{cox00}, the 
 total gas mass including other elements, such as helium, becomes $M 
 _{\rm gas}=1.41\,M\rm _{H_2}$.  
The surface mass density of molecular hydrogen on the galaxy plane is 
calculated from the integrated CO-line intensity, $I_{\rm CO}$, 
the galaxy inclination, $i$, and the conversion factor, $X_{\rm CO}$, as 
\begin{eqnarray} 
\left(\frac{\Sigma_{\rm H_2}}{\: M_\odot\,\rm pc^{-2}}\right)=&&3.0\times 10^2\cos 
 i\left(\frac{I_{\rm CO}}{\rm Jy\, km\, 
 s^{-1}\,arcsec^{-2}}\right)\nonumber \\
&&\times\left[\frac{X_{\rm CO}}{1.8\times 
 10^{20}\rm cm^{-2}\,(K\,km\,s^{-1})^{-1}}\right]. 
\end{eqnarray} We adopt the conversion factor 
of $S_{\rm CO}\equiv N_{\rm H_2}/I_{\rm CO}=1.8\times 10^{20}\rm cm^{-2}(K\, km 
s^{-1})^{-1}$ \citep{DHT01} from observations in 
the Galaxy, while $X_{\rm CO}$ could be smaller in galactic center 
regions \citep{AST96}. The total flux within 
$r=5\: \rm kpc$ ($32''$) is $S_{\rm CO}=3.9\times 
 10^2\rm Jy\,\: \rm km\,s^{-1}$, which 
corresponds to $M_{\rm gas}=1.0\times 10^9 M_\odot$ for the galaxy distance 
$D=16.1\; \rm Mpc$. The dynamical mass is derived from rotation curve 
as $M_{\rm dyn}=rv^2/G$, assuming circular rotation. The ratio of the dynamical mass, 
$4.3\times 10^{10} \: M_\odot$ ($v=270\: \rm km\,s^{-1}$), to the total molecular 
gas mass, $M_{\rm gas}/\it M_{\rm dyn}$, is about 2.4\% within $r\sim 5\: \rm kpc$. 
 
\begin{figure}[t]
\begin{center} 
\FigureFile(80mm,80mm){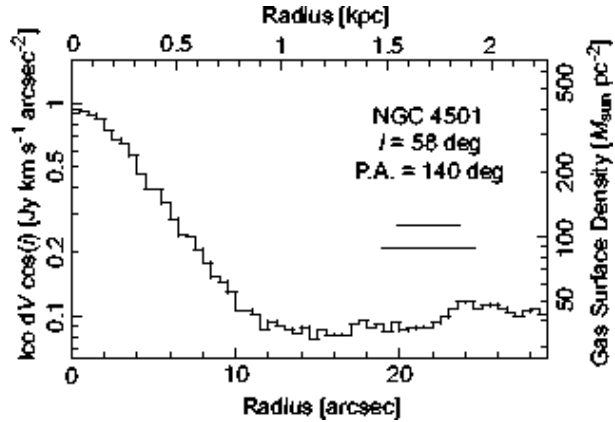} 
\end{center} 
\caption{Radial distribution of the molecular gas surface density in NGC 
 4501 with a logarithmic scale. The CO intensities were corrected for 
 primary beam attenuation 
 and inclination, and averaged over annuli 
 with $\timeform{0''.5}=39.0\: \rm pc$. $X_{\rm CO}=1.8\times 10^{20}\rm cm^{-2} 
(K\: \rm km\,s^{-1})^{-1}$ 
 and $\Sigma_{\rm gas}=1.41\Sigma_{\rm H_2}$ was adopted to calculate the 
 total gas surface density in unit of mass per parsec square. The two 
 horizontal bars show the beam size along the major and minor axes 
 (\timeform{5''.6}$\times$ \timeform{3''.7}).}   
\label{fig:rad} 
\end{figure} 
\subsection{Spiral Arms} 
Two molecular spiral arms extend out from the center to the end of our 
field of view in the map. The arms start from the northern side and the 
southern side of the nucleus, and continue to the northeast arm 
and the southwest arm, respectively.  
Their outer parts are less clear owing to the 
primary beam attenuation. These spiral arms 
are distinct up to $r\sim 50''$ in larger spatial coverage, but lower 
resolution map by \citet{WB02}. Our result shows 
these arms more clearly, owing to the higher resolution ($\sim 5''$).  
It also shows many giant molecular cloud associations in the arms.  
These arms are coincident with the dust lanes ($r\sim 2.9\: \rm kpc$) in 
 an optical image (figure \ref{fig:B}). The dust lanes 
  are continuous down to the nuclear region of $\sim 500\: \rm pc$. 
The line-of-sight velocities along these arms  
show deviations from that of a circular-rotating disk.  
It is about $\sim 50\: \rm km\,s^{-1}$ on the 
galactic minor axis. They appear in the 
position--velocity diagram as a ``figure-of-eight'' pattern (figure \ref{fig:pv}).  
These distortions  
in the velocity field is similar to that of the H\emissiontype{I} spiral 
arms of M 81 \citep{Visser80}, which are typical in gaseous spiral 
arms caused by density waves. 
\subsection{Central Double Peaks} 
\begin{figure*}[t] 
\begin{center} 
\FigureFile(160mm,100mm){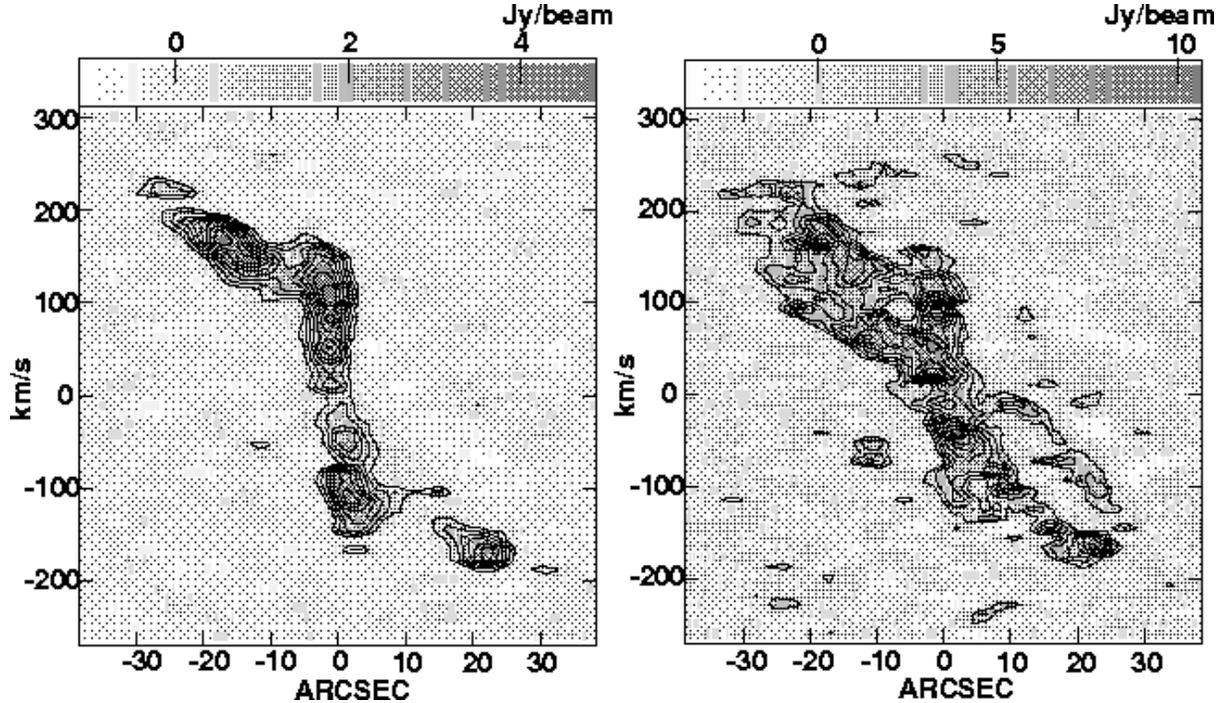} 
\end{center} 
\caption{Low-resolution position--velocity diagram along the major axis 
 of NGC 4501. Velocities are not corrected for inclination ($58^\circ$). Left-hand side corresponds to 
 the northwest. Contour levels are 20, 30, 40, 50, 60, 70, 80, 90\% of 
 the peak value. Left: The slit width is central $5''$. The peak 
 value is $4.9 \, \rm Jy\, beam^{-1}$ or 51 K. The  
 central steep rise is resolved into double peaks. Right: The 
 slit width is $60''$, including extended spiral arms. The peak value is  
$11\, \rm Jy\, beam^{-1}$ or 110 K. The  
 figure-of-eight pattern due to the spiral arms are apparent. 
 } 
\label{fig:pv} 
\end{figure*}

\begin{figure*}[t]
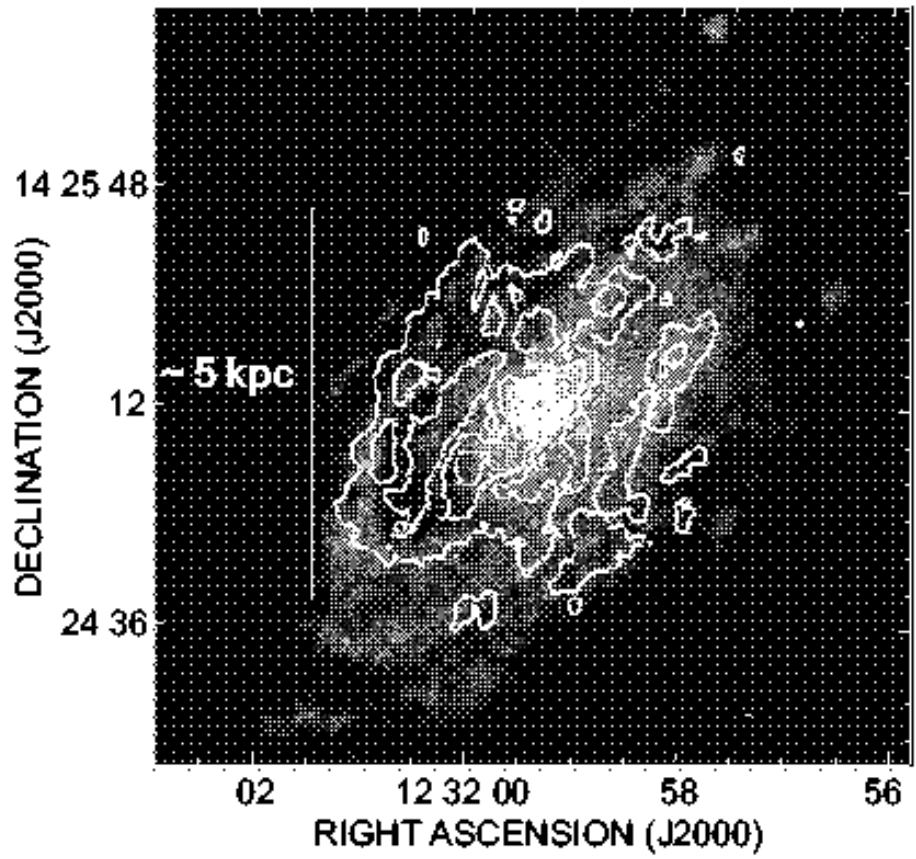

\begin{center} 
\FigureFile(120mm,120mm){figure6.epsi} 
\end{center} 
\caption{Comparison of the low-resolution CO map (contours) and the {\it
 B}-band image (gray scale, from the DSS archive). Northeastern CO 
 spiral arms trace dark dust lanes.} 
\label{fig:B} 
\end{figure*}

The low-resolution map shows a central condensation of 
radius $\sim 5''$($\sim 390\: \rm pc$). The total CO flux of this component is 
$S_{\rm CO}=48\, \rm Jy\: \rm km\,s^{-1}$, which 
corresponds to $M_{\rm gas}=1.3\times 10^8 \it M_\odot$. 
The gas-to-dynamical mass ratio within $r=5''$ ($390\: \rm pc$) is $\sim 
3.5\%$, since the dynamical mass derived from $v\sim 200\: \rm km\,s^{-1}$ at $r=5''$ is  
$M\rm_{dyn}=3.6\times 10^9 \it M_\odot$. The low-resolution ($5''$)
velocity field indicates a slight non-circular motion. 
 This nuclear component shows 
high velocities of $\sim\pm 200\: \rm km\,s^{-1}$ in the 
position--velocity diagram, and is resolved into 
two bright peaks (figure \ref{fig:pv}) with distinct velocities.  
 
The nuclear concentration is also resolved into double peaks  
 in the high resolution ($\sim 2''$) intensity map, separated by 
$\timeform{4''.7}$ ($\sim 370\: \rm pc$) (figure \ref{fig:high}). 
 Each of these corresponds to the 
 double peaks in the P--V diagram. The net flux within $r\sim 5''$ is 
 $14\:\rm Jy\: \rm km\,s^{-1}$ in high resolution, while it is $48\:\rm Jy\: \rm km\,s^{-1}$ in low 
 resolution, because the higher resolution map missed the extended fluxes.  
The double peaks may be on more diffused components, which are unseen in 
 our image. 

The peak intensity in the high-resolution map is $\rm 9.6\, Jy\, 
 beam^{-1}\, km\, s^{-1}$ or 290 $\rm K\, km\, s^{-1}$, which 
 corresponds to ${\rm \Sigma_{\rm H_2}}=440\: M_\odot {\rm pc^{-2}}$ and  
${\rm \Sigma_{\rm gas}}=620\: M_\odot {\rm pc^{-2}}$.  
\section{Gas Dynamics in NGC 4501} 
As we have shown above, the observed velocity field demonstrates 
distinct non-circular motions along the spiral arms. Its well-ordered 
pattern indicates that the density wave, rather than the stochastic 
process, dominates in the central region of this galaxy.  
Although this galaxy seems 
to be multiple-armed in the outer region, there is some evidence that it has 
two arms in the central $r\lesssim 40''$.   
\citet{Elmegreen99} showed a continuous two-armed spiral in the
$K'$-band image, and 
concluded that in this galaxy there are density waves caused by the 
spiral arms of an older stellar population. An unsharp-masked 2MASS 
{\it K}-band image is shown as figure \ref{fig:K}.  
 
Optically unbarred galaxies sometimes unveil their hidden bars in the  
images at longer wavelengths (e.g., Scoville et 
al. 1988; \cite{MulchaeyRK97}; \cite{kohno03}), because contributions 
from young stars and disturbances by dust lanes may make a spiral-like 
appearance at shorter wavelengths. 
Several studies have examined this galaxy for a bar, by means of 
elliptical isophote fitting on optical and near-infrared images 
(\cite{Jungwiert97}; \cite{Elmegreen99}; 
\cite{Sil'chenko99}). 
 If a bar structure is dominant, the position angles, and 
eccentricity of the isophotal ellipses are expected to  
vary with the radius, since the bars are triaxial and their eccentricity shows a 
radial  
variation. Near-infrared images are preferable for this purpose because 
of their advantage to reflect a galactic potential better than optical images.  
The criteria for bar identification in general is: (1) the ellipticity 
increases as a function of the radius, and then decreases to 
reveal the inclination of the disk; (2) the position angle is constant 
over the radii where the ellipticity is rising (\cite{MR95}; 
 \cite{MulchaeyRK97}). 
\citet{Jungwiert97} and \citet{Elmegreen99} independently conducted this analysis 
in the {\it H}-band and the $K'$-band, respectively, and obtained quantitatively 
almost the same results. Interior to $r\sim 15''$, position angle 
remains constant at about $140\degree$, and the ellipticity 
gradually rises from zero to $\sim 0.4$. Then, P.A. declines to reach the 
minimum value of
$\sim 120\degree$ at $r\sim 27''$, and gradually recovers to attain its 
original value of $140\degree$ at $r\sim 50''$, and then keeps this 
value constantly to the outer region. The ellipticity also drops at $r\sim 
20''$ to 0.3, and then gradually increases to 0.5 at $r\sim 40''$.  
(see figure 4 in Jungwiert et al. and figure 5 in Elmegreen et al.) 
 
From their quantitatively same results, Jungwiert et al. concluded 
that there is no evidence for a bar or other 
triaxiality, while Elmegreen et al. concluded the opposite. 
Giving a certain criteria for presence of a bar, and based on their 
careful simulations of the effect of projection on an identification of 
a bar, Jungwiert et al. concluded that the change of ellipticity in the 
inner region and the constancy of the P.A., as well as the fact that its value 
corresponds to that of the outer disk, is compatible with a spheroidal 
shape of the bulge. The slight change of the eccentricity and P.A. behind $r\sim 
18''$ should be attributed to spiral arms.
\onecolumn
\begin{figure*}[b]
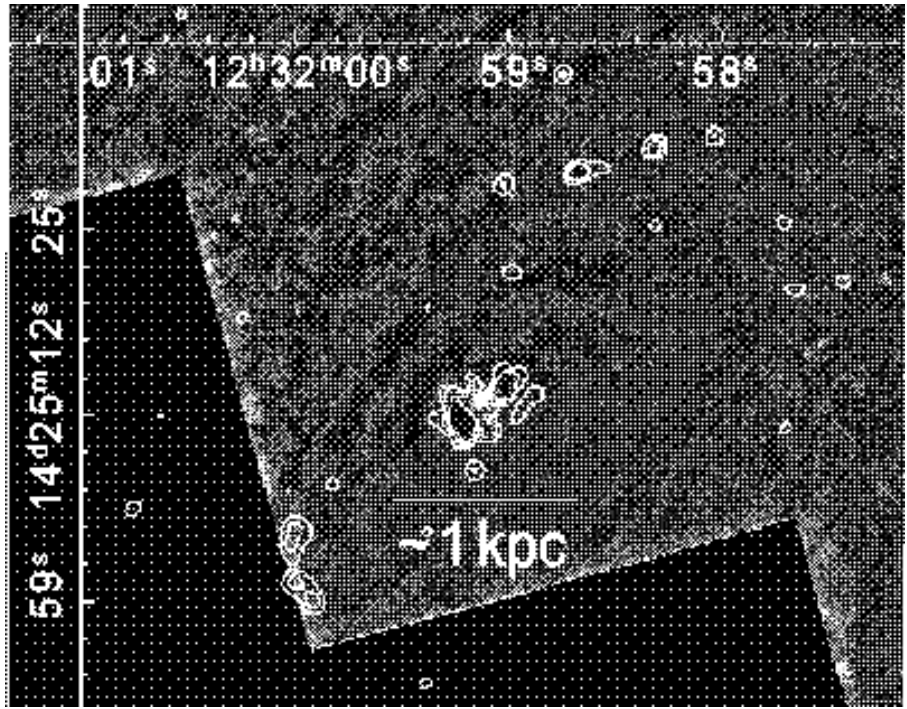
 
\begin{center} 
\FigureFile(120mm,120mm){figure7.epsi} 
\end{center} 
\caption{Comparison of a high-resolution CO map (contours) and 
an unsharp-masked HST 
 wide $V$-band image (gray scale). CO double peaks are 
 located on the roots of spiral dust lanes.} 
\label{fig:HST} 
\end{figure*} 
\twocolumn
$\!\!\!\!\!\!\!\!\!\!\!$
On the other hand, Elmegreen et al. noted the presence of a small 
bar of $15''$ diameter (see their table 1), without any explanation about the 
criteria to identify the bar. It seems that they might judge from the 
growing ellipticity in the inner region, however, as discussed in  
Jungwiert et al.(1997), that a bar is not relevant to explain those characteristics  
in this region. 
 
Moreover, no large-scale bar has been found in NGC 4501, and  
the HST $V$-band image shows the 
spiral arms and spiral dust lanes penetrating into the nucleus (figure \ref{fig:HST}).  
Based on these facts, we consider the spiral structure 
to be prominent, rather than a bar, in the gas dynamics of NGC 4501. 
 
In the following section, we calculate the gas cloud orbits in a 
potential made by stellar spiral arms by modifying the damped-orbit 
model of \citet{wada94} to understand the gas motion in NGC 
4501. Based on the spiral-arm driven gas dynamics, we suggest 
mechanisms of the angular-momentum transfer in spiral potentials, which 
could produce the central gas condensations, and probably, the double 
peaks. 

\begin{figure}[t]
\begin{center} 
\FigureFile(80mm,80mm){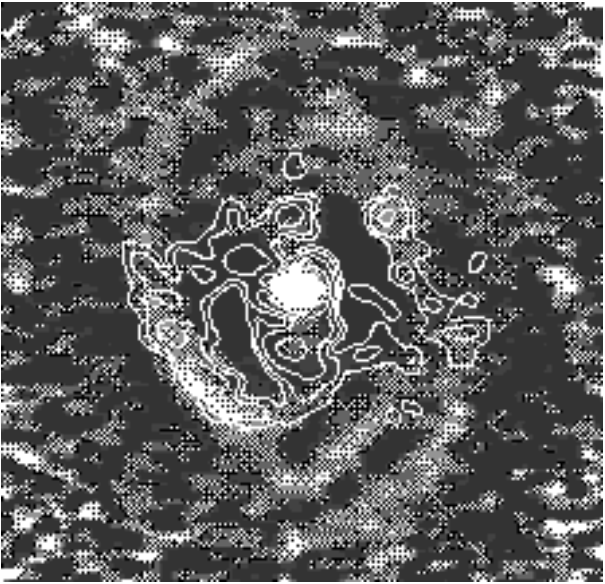} 
\end{center} 
\caption{Deprojected overlay of an unsharp-masked {\it K}-band image 
 (gray: from the 2MASS archive). The vertical axis is taken to be 
 galactic major axis. } 
\label{fig:K} 
\end{figure} 
 
\begin{figure}[h]
\begin{center} 
\FigureFile(80mm,80mm){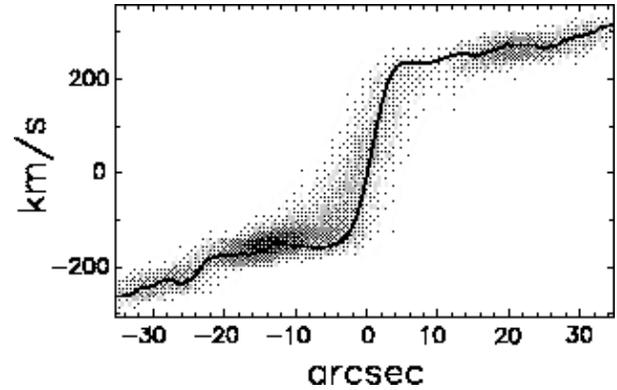} 
\end{center} 
\caption{Rotation curve derived by an iteration method of 
 Takamiya and Sofue (2002) and reproduced position--velocity diagram by 
 convolving the rotation curves with the observed intensity distribution. 
The velocities are corrected for inclination.} 
\label{fig:rc} 
\end{figure} 
\subsection{Damped-Orbit Model in Spiral Potential} 
\begin{figure} 
\begin{center} 
\FigureFile(80mm,80mm){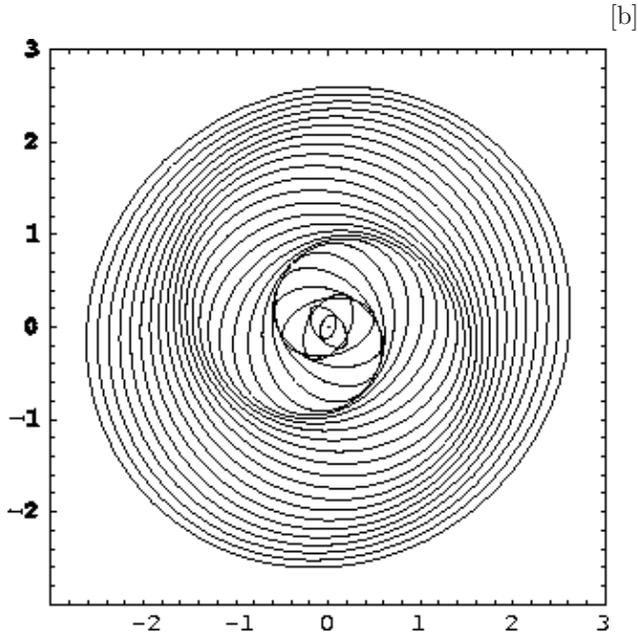}[b] 
\end{center} 
\caption{Particle orbits in the fixed spiral 
 potential. An orbit-crowding region is generated as a spiral pattern. Used 
 parameters: $\theta=17\degree$, 
 $\epsilon=0.03$, $\Lambda\equiv\lambda /\kappa =0.1$, $\Omega \rm _b=0.01$,  
$a=1$. } 
\label{fig:orbit} 
\end{figure}

In order to study the gas responses in a weak spiral potential, we 
modified the damped-orbit model for a bar potential \citep{wada94}. This 
model gives closed gas orbits in a barred potential by analytically  
solving the 
equations of motion, which include a damping-force term,  
to emulate the collisional nature of the gas.  
 It has been applied to galaxies NGC 5005 \citep{Sakamoto00} 
 and NGC 3079 \citep{Koda02}, and successfully 
explains the observed gas motions in these barred galaxies. 
Instead of a straight bar potential, we used a spiral 
potential.  
Modeling after \citet{BT87},  
we assume that the stellar potential is represented by an  
axisymmetric disk potential, $\Phi_0$, and a first-order perturbation 
owing to the spiral arms, $\Phi_1$, as  
\begin{equation} 
\Phi(R,\psi)=\Phi_0(R)+\Phi_1(R,\psi). 
\end{equation} 
We use the Plummer potential for the axisymmetric disk part, 
\begin{equation} 
\Phi_0(R)=-\frac1{\sqrt{R^2+a^2}}. 
\end{equation} 
The 
 scale length of the model is normalized by $a$. 
 $\Phi_1$ is the first-order perturbation by spiral arms, written as 
\begin{equation} 
\Phi_1(R,\psi)=\Phi_{\rm p}(R)\cos [m(\Omega_0-\Omega_{\rm p})t+p\log(R)], \label{plummer} 
\end{equation} 
 which 
represents a logarithmic $m$-armed spiral with $p\equiv m\cot \theta$, 
where $\theta$ is the pitch angle of the spiral arms. We set $m=2$ and 
$\theta=17\degree$ to represent the {\it K}-band 
feature of NGC 4501.  
The amplitude of the deviation from the axisymmetric part is taken to be 
\begin{equation} \Phi_{\rm p}(R)=\epsilon\frac{aR^2}{(R^2+a^2)^2}, \label{sanders} 
\end{equation} 
which was used for bar models by \citet{Sanders77} and \citet{wada94}. 
We solve the equations of motion for test gas particles moving in this 
potential (see appendix 1 for the detailed 
description for damped-orbit model in a spiral potential). 
 The resultant orbits are shown in 
figure \ref{fig:orbit}. Oval orbits appear, whose major axes change with
the radius. As a result, spiral-shaped orbit-crowding regions are 
generated.  
 
\subsection{Comparison with CO Distribution and Velocity Field} 
Figure \ref{fig:model} shows the density distribution and velocity field obtained from 
orbit calculations. They are projected with the 
inclination and position angle of NGC 4501. The density distribution was 
simply estimated based on the assumption that the staying time of a gas cloud 
at each point is 
inversely proportional to the velocity. The observed radial profile was
applied. After that, we convolved the model images with a Gaussian 
beam of FWHP $65''$ to reflect the primary-beam attenuation of NMA. 
Our model (figure \ref{fig:orbit}) shows that the 
clouds slow down near to the major axes of the oval orbits, which make  
the orbit-crowding region. 
The cloud distribution becomes similar to the orbit-crowding 
pattern, and gas clouds moving in a spiral potential gather along the 
spiral pattern. This naturally explains the observed molecular 
arms.  
The distortions along the spiral arms 
and in the nuclear region of the model velocity field well represent that of 
the observations.  
The velocity field in figure \ref{fig:model} shows that the trends in
the line-of-sight 
velocity of gas clouds change across the arms. On the far side of the arms to 
 us, gas is approaching compared to circular motion. When we look nearer 
 and nearer, its line-of-sight velocity approaches that of the circular 
 motion. After that, it turns to receding motion, and gradually 
 returns to the approaching motion. This behavior is well reproduced by 
 this model.  
\section{Discussion} 
\subsection{Central Double Peaks} 
Figure \ref{fig:HST} shows that 
the double peaks inhabit the root of spiral dust lanes penetrating into 
the nucleus. The nuclear spiral dust lanes have recently been found 
 in many galaxies, and are thought to be the location of shock, 
 and trace the sites of angular-momentum 
 transfer, since shocks cause a loss of kinetic energy.
 (e.g. \cite{regan99}; \cite{martini03}).   
The CO ``twin peaks'' are often found in barred galaxies 
 (e.g. \cite{Kenney92}). The ``twin peaks'' are connected to large-scale dust lanes, 
 and regularly  
 accompany to nuclear star-forming rings. They show a gas-to-dynamical 
 mass ratio of as high as $\sim 0.3$, and are thought to be 
 gravitationally unstable enough for star formation. They are often 
 considered to be a  
 the orbit-crowding region with an $x_1/x_2$ orbit population change 
near to the 
 inner Lindblad resonance.  
 On the other hand, the gas-to-dynamical mass ratio within $r=5''$ 
 ($390\: \rm pc$) is $\sim 3.5\%$ and no nuclear star-forming ring is seen 
 near to the double peaks in NGC 4501  
 (see figure 5 in \cite{koopmann01}). This is 
 the first time that this kind of ``twin peaks'' was found in a non-barred 
 galaxy, although the physical conditions are significantly different from 
 that of the ``twin peaks'' in barred galaxies.  
An inner Lindblad resonance also exists in the spiral potential, and  
the appearance of double peaks might indicate  
that they are associated with the
 inner Lindblad resonance, as well as those in barred galaxies. 
 However, it cannot be judged from our 
 present data only, because the central rotation curves are derived from
 a $5''$ 
 resolution velocity field, while the double peaks are at  
$r\sim\timeform{2''.4}$. In fact, the epicyclic frequency, 
 $\kappa\equiv\sqrt{(d^2\Phi_0/dR^2)+3\Omega^2}$, is 
 sensitive even to small errors in the rotation curves. Observations at higher 
 resolution and sensitivity still remain as future work.   
\subsection{Central Condensation and Mechanisms of Angular Momentum Transfer} 
 As revealed by Sakamoto et al. (1999a), central CO condensations within
   $r\sim 500\: \rm pc$ are prevalent in CO-luminous galaxies, and NGC 4501 is one 
 of those with central CO excess. In terms of its radial profile, 
 the much shorter {\it e}-folding radius ($\timeform{4''.3} \sim 340\: \rm pc$; 
 this work) than that of the global CO disk ($\timeform{47''.0} \sim 
 3.8\: \rm kpc$; \cite{nishiyama01}) evidences some distinct origins of the  
central gas disk from that of the global gas disk. Thus, it indicates a 
possible radial inflow of the gas toward the central disk.  
 
There are two possible mechanisms to transfer 
angular momentum and transport the gas into the central region {in a 
global perturbation, like bars or spirals. In the following, we 
consider these mechanisms in a spiral potential, and estimate the 
effectiveness.\\ 
(1) \it Galactic Shock Wave:\rm\, The 
calculation shows that the  
orbits are crowded along the spiral pattern. It is likely that 
galactic shock occurs in these orbit-crowding regions, and that this leads 
to the loss of angular momentum of the gas clouds. We estimated this loss by  
considering the oblique shock of isothermal gas. A detailed description 
is shown in appendix 2. The strength of the shock 
depends on the angle between the shock front and the gas streamline. 
With simple approximations, the loss of angular momentum is 
$\sim \sin^2\theta$ times the initial value after a shock. Adopting 
a pitch angle of $\theta=17\degree$ for NGC 4501 derived from 
$K'$-band image, the loss is estimated to be 16\% 
after an orbital rotation.  
 \\ 
(2) \it{Drag of Stellar Spiral Potential}: \rm Gas clouds show 
elliptical orbits in a nonaxisymmetric potential made by stellar spiral 
arms. Therefore, when a cloud is in the leading side of the potential valley, gravitational torques 
from the stellar spirals are exerted to drag the cloud `backwards',
which result 
in angular-momentum transfer from the gas to the stars. 
Note that gas clouds will oppositely obtain angular 
momentum when they are on the trailing side. Whether a gas cloud  
 loses or obtains angular momentum 
\onecolumn
\begin{figure*}[b]
\begin{center} 
\FigureFile(160mm,320mm){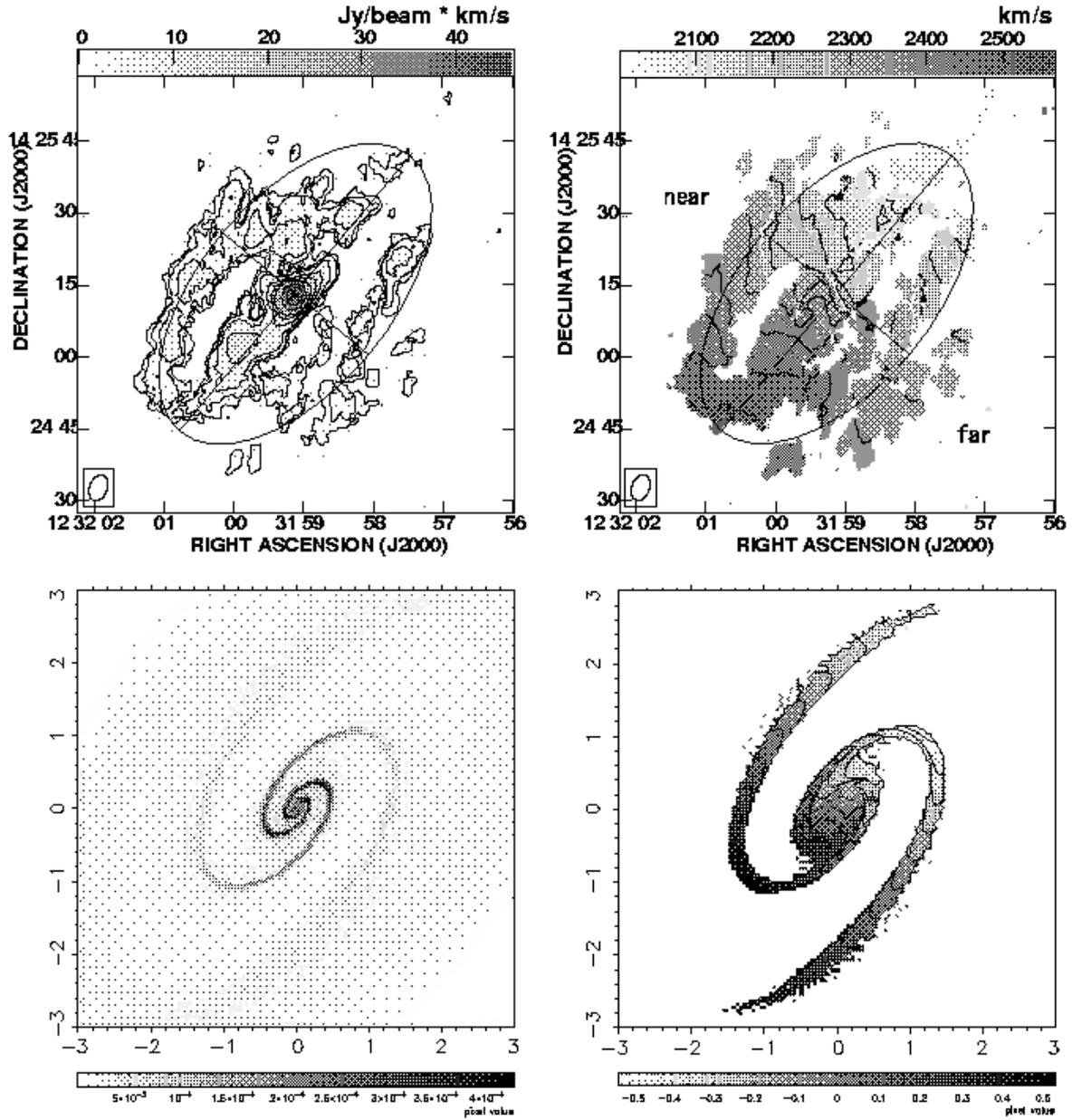} 
\end{center} 
\caption{Gas distributions (left panels) and velocity fields (right 
 panels) obtained by observations 
 (upper panels) compared to those by our models (lower panels).  
The parameters used for 
 the model are the same as that used for figure 10. The model disk is 
 inclined with the inclination and position angle of NGC 4501.} 
\label{fig:model} 
\end{figure*} 
\twocolumn
$\!\!\!\!\!\!\!\!\!$ at a certain  
radius is determined by integration of this 
torque along the orbit of the gas, as shown in equation (\ref{torque}). 
 In a bar potential,  
the phase of the major axis of 
elliptical orbits always leads the bar potential inside of the corotation 
resonance (CR); therefore, clouds stay longer in the leading side of the 
potential valley  
than in the trailing side. This leads to a loss of the angular momentum and 
inflow of the gas inside CR (Wada 1994). 
We calculated the change of angular momentum, $\Delta \dot{J}_z(R_0)$, for 
a spiral potential of $\theta=17\degree$, and obtained the loss of the 
angular momentum, $\Delta 
\dot{J}_z(R_0)/J_z(R_0)=0.8\%$, at most, using the parameters presented with 
figure \ref{fig:orbit}. This value depends on spiral perturbation strength, 
$\epsilon$, and pitch angle, $\theta$. 
 
The above estimations show that the angular-momentum variation due to galactic 
shock (16\%) is an order-of-magnitude larger than that of gravitational 
torques (0.8\%). 
Therefore, gas clouds are destined to lose angular momentum due to 
 shocks regardless of a gain or loss due to the torques from the 
stellar spiral potential. 
\section{Conclusions} 
We have observed CO(1--0) emission from the non-barred Seyfert 2 galaxy 
NGC 4501 to study the gas dynamics in the central 5 kpc region. 
 
1. In the central 5 kpc, NGC 4501 has two molecular gas 
components --- spiral arms penetrating into the nuclear region, and a central 
concentration with radius $5'' (\sim 390 \: \rm pc)$, and the gas mass $M_{\rm 
gas}=1.3\times 10^8 \it \: M_\odot$. 
 
2. The spiral arms are observed from the nuclear region to the end of 
 our field-of-view ($r\sim 40''$). 
 The velocity field along these arms shows deviations from circular 
 motion. That is about $\sim 
 50\: \rm km\,s^{-1}$ in line-of-sight velocity on the galactic minor axis. 
 
3. The central concentration is resolved into double peaks separated 
by $\timeform{4''.7} (\sim 370\: \rm pc)$ in the high-resolution map. 
 The double peaks are located on the root of nuclear dust spirals 
 found by HST. 
  
4. The well-ordered non-circular motion reveals that 
the gas is driven by density waves in the central region of this galaxy, rather than by stochastic processes.  
This result is consistent with that inferred from near-infrared observations.  
 
5. To understand gas motions in NGC 4501, we studied the 
gas cloud orbits in spiral potential with the damped-orbit model. The 
result well explains the velocity field. 
The gas dynamics in NGC 4501 can be well understood
 as being governed by a spiral potential.  
The observed density distribution can also be explained by a simple 
assumption that the staying time of clouds is inversely proportional to 
its velocity at each point.  
  
6. Centrally-condensed double peaks have a low 
star-forming efficiency, which may come from gravitational stability of 
gas, owing to low gas-to-dynamical mass ratio. It may be possible that  
they are the results of angular-momentum transfer due to shocks 
 in nuclear dust spirals.  
 
7.  
Based on our model, we estimated the loss of angular momentum in 
NGC 4501 due to two possible mechanisms --- galactic shock in 
orbit-crowding regions and  
gravitational torques exerted by the stellar spiral potential. 
The effect of galactic shock is an order-of-magnitude larger than 
that of gravitational torques. 
Stellar spiral arms are responsible for the 
angular-momentum loss of gas, and may lead to gas inflow toward the galactic 
centers. The gas inflow rate depends on the 
galactic morphology. Our spiral model and its consequential gas inflow 
are natural extensions of those discussed for a bar.  
The spiral-shock inflow mechanism, as well as a bar-induced inflow, 
 can be a potent 
instrument to study the relation between the galactic morphology and 
the central gas features, such as nuclear condensations and starbursts. 
\\ 
 
The authors thank an anonymous referee for comments, which refined
this paper. We are indebted to the NRO staff for their support 
with observations. We are grateful to Hiroyuki Nakanishi and Tsutomu 
 Takamiya for 
their help with observations and reductions and for useful comments. 
We also appreciate 
Keiichi Wada for his profitable comments.   
S. O. and J. K. are financially supported by the Japan Society for the Promotion of Science 
(JSPS) for Young Scientists.  
\section*{Appendix 1. Gaseous Orbits in Spiral Potential} 
In section 4, we used the damped orbit model to describe gas 
motions in a spiral potential. This model follows \citet{BT87}, 
which described analytic solutions of equations of motion  
for collisionless particles in a weak bar, and obtained stellar orbits 
under an epicyclic approximation. 
Wada (1994) introduced the damping force term considering the  
collisional nature of gas, and obtained gas orbits in a bar potential. 
We modify this damped-orbit model to 
introduce a spiral potential, instead of a barred potential.  
We will consider the gas clouds moving in a fixed stellar potential. The 
self gravity of gas is not 
taken into account.  
 
We assume that the pattern of a potential rotates with a certain pattern 
speed, $\Omega_{\rm p}$. Let ($R, \psi$) be polar coordinates in a frame that 
rotates with the potential, $\Phi (R,\psi)$.  
The equations of motion of a test particle in this frame are written as 

\begin{equation} 
\ddot{R}-R\dot{\psi}^2=-\frac{\partial\Phi}{\partial 
R}+2R\dot{\psi}\Omega_{\rm p}+\Omega_{\rm p}^2R,  
\tag{A1}  
\label{eqm1} 
\end{equation} 

\begin{equation} 
R\ddot{\psi}+2\dot{R}\dot{\psi}=-\frac1R\frac{\partial\Phi}{\partial\psi} 
-2\dot{R}\Omega_{\rm p}.  
\tag{A2} 
\label{eqm2} 
\end{equation} 

Here, we derive analytic solutions of equations of motion to 
the first order. We assume that the non-axisymmetric part of the 
potential, $\Phi_1$,
is much weaker than the axisymmetric part, $\Phi_0$, then  

\begin{equation} 
\Phi(R,\psi)=\Phi_0(R)+\Phi_1(R,\psi), 
\tag{A3} 
\end{equation} 

where $|\Phi_1 /\Phi_0| \ll 1$. We divide $R$ and $\psi$ into zeroth- and 
first-order parts:

\begin{equation} 
R(t)=R_0+R_1(t);\quad \quad \psi(t) =\psi_0(t)+\psi_1(t). 
\tag{A4} 
\end{equation} 

We then substitute these expressions into equations (\ref{eqm1}) and 
(\ref{eqm2}).  
Now, the first-order terms become 

\begin{equation}  
\ddot{R}_1+\left[\frac{d^2\Phi_0}{dR^2}-\Omega^2\right]_{R_0} 
R_1-2R_0\Omega_0\dot{\psi}_1=-\left[\frac{\partial\Phi_1}{\partial R}\right] 
_{R_0},  
\tag{A5} 
\label{eqm3} 
\end{equation} 

\begin{equation} 
\ddot{\psi}_1+2\Omega_0\frac{\dot{R}_1}{R_0}= 
-\frac{1}{R_0^2}\left[\frac{\partial\Phi_1}{\partial\psi}\right]_{R_0}, 
\tag{A6} 
\label{eqm4} 
\end{equation} 

where  

\begin{equation}
\Omega(R)\equiv\pm\sqrt{\frac1R\frac{d\Phi_0}{dR}}, 
\tag{A7}
\end{equation}

\begin{equation}
\Omega_0\equiv\Omega(R_0)\quad ;\quad \psi_0\equiv(\Omega_0-\Omega_{\rm p})t,
\tag{A8}
\end{equation}

\begin{equation}
\kappa_0\equiv\sqrt{\frac{d^2\Phi_0}{dR^2}+3\Omega^2}.
\tag{A9}
\end{equation}

We now introduce a spiral-shaped perturbation, 
\begin{equation} 
\Phi_1(R,\psi)=\Phi_{\rm p}(R)\cos[m(\Omega_0-\Omega_{\rm p})t+p\log(R)], 
\tag{A10} 
\end{equation} 
which represents an $m$-armed logarithmic spiral with $p\equiv m\cot\theta$, 
where $\theta$ is the pitch angle of the spiral arms.  
Next, we add a damping-force term to the equation of 
motion (\ref{eqm3}) following Wada (1994). The damping force term represents 
the effects of the collisional nature of gas; thus, a plausible approximation 
can be the form of $2\lambda \dot{R}_1$, which is proportional to the 
velocity of the radial oscillation, and an appropriate value, $\lambda$. 
 
Then, the equations of motion become 
\begin{align*} 
\ddot{R}_1+2\lambda\dot{R}_1+\kappa_0^2R_1=&f_0\cos[m(\Omega_0-\Omega_{\rm p})t 
+\alpha]\\
&+f_1\sin[m(\Omega_0-\Omega_{\rm p})t+\alpha], 
\tag{A11} 
\end{align*} 

\begin{align*} 
\ddot{\psi}_1+2\Omega_0\frac{\dot{R}_1}{R_0} 
=&-\left[\frac{\partial\Phi_{\rm p}(R)}{\partial R}\right]_{R_0} 
\cos\left[m(\Omega_0-\Omega_{\rm p})t+\alpha\right]\\
&+\Phi_{\rm p}(R)\frac{C_1}{R_0} 
\sin\left[m(\Omega_0-\Omega_{\rm p})t+\alpha\right]. 
\tag{A12} 
\end{align*} 

We assume the axisymmetric part of the potential, $\Phi_0(R)$ to be
Plummer potential (\ref{plummer}), and  
the strength of the non-axisymmetric part, $\Phi_{\rm p}(R)$, to be
equation (\ref{sanders}),  
following Sanders (1977) and Wada (1994). 
When we assume the gas orbits to be closed, 
the solutions for these equations are 

\begin{align*} 
R_1=&\frac{1}{C_1^2+C_2^2}
\left\{(C_1f_0-C_2f_1) \cos[m(\Omega_0-\Omega_{\rm p})t+\alpha] \right. \\
&\left. +(C_2f_0+C_1f_1) 
\sin[m(\Omega_0-\Omega_{\rm p})t+\alpha]\right\}, 
\tag{A13} 
\end{align*} 

\begin{align*} 
\psi_1=-&\frac{1}{C_1^2+C_2^2} 
\frac{2\Omega_0}{mR_0(\Omega_0-\Omega_{\rm p})}
\\ 
&\times \left\{ (C_1f_0-C_2f_1) \sin[m(\Omega_0-\Omega_{\rm p})t+\alpha]\right.\\  
&\left. \quad -(C_2f_0+C_1f_1)\cos[m(\Omega_0-\Omega_{\rm p})t+\alpha] \right\}\\
-&\frac{\Phi_{\rm p}(R)}{mR_0^2(\Omega_0-\Omega_{\rm p})^2} 
\sin[m(\Omega_0-\Omega_{\rm p})t+\alpha], 
\tag{A14}
\end{align*} 

where 

\begin{equation}
\alpha\equiv p\log(R_0),
\tag{A15}
\end{equation}

\begin{equation}
f_0\equiv 
-\left[\frac{d\Phi_{\rm p}}{dR}+\frac{2\Omega\Phi_{\rm p}} 
{R(\Omega-\Omega_{\rm p})}\right]_{R_0},
\tag{A16}
\end{equation}

\begin{equation}
f_1\equiv\Phi_{\rm p}(R_0)\left[\frac{\partial\alpha(R)}{\partial 
R}\right]_{R_0},
\tag{A17}
\end{equation}

\begin{equation}
C_1=\kappa_0^2-m^2(\Omega_0-\Omega_{\rm p})^2\: ,\quad C_2=2\lambda 
m(\Omega_0-\Omega_{\rm p}).
\tag{A18}
\end{equation}

 Particles trace oval orbits whose position angle of the major axes  
changes continuously due to both the damping term and the spiral potential.  
This situation is different from that of stellar orbits in a bar potential,  
whose major axes can only be parallel or perpendicular to the bar.   
\section*{Appendix 2. Estimation of Angular-Momentum Variation} 
\subsection*{A2.1. Galactic Shocks} 
Behaving as a fluid, interstellar gas becomes supersonic by  
a variation of the local gravitational field due to the density wave  
(\cite{Fujimoto66}; \cite{Roberts69}). This  
results in a galactic shock wave. It is possible that gas loses angular 
momentum due to galactic shocks along the spiral arms.  
We estimate the effect of the shock with simple 
approximations of isothermal gas with a Mach number of 
$M\gg 1$. Although the orbits are non-circular, a circular-orbit 
approximation is enough for an order-of-magnitude estimation.  
The Mach number, $M$, is defined as $M=v/a$ with velocity 
$v$ and sound speed $a$. For interstellar gas, because $M\sim 10$ with $v\sim 
100\: \rm km\,s^{-1}$ and $a\sim 10\: \rm km\,s^{-1}$, $M\gg 1$ is plausible. 
 
We now consider the gas bumping into a shock front with velocity $v_1$ and 
angle $\alpha$. In the jump condition of isothermal 
oblique shock, the velocity perpendicular to the shock front is reduced 
by $1/M^2$. It leaves then there with velocity $v_2$ and angle 
$\beta$, which are written as 
\begin{equation} 
v_2=v_1\sqrt{\frac{\sin^2\alpha}{M^4}+\cos^2\alpha}, 
\tag{A19} 
\end{equation} 
\begin{equation} 
\beta=\tan^{-1}\left(\frac{\tan\alpha}{M^2}\right). 
\tag{A20} 
\end{equation} 
Thus the change in the angular momentum, $J_1-J_2$, becomes 
\begin{align*} 
J_1-J_2&= rv_1\cos(\theta-\alpha)-rv_2\cos(\theta-\beta) \\ 
    &= rv_1\left[\cos(\theta-\alpha)-\sqrt{1+\frac{\tan^2\alpha}{M^4}}\, \right.\\ 
&\qquad \left. \times\cos\alpha\,\cos\left(\theta-\tan^{-1}\frac{\tan\alpha}{M^2}\right)\right]. 
\tag{A21} 
\end{align*} 
Then, the supersonic condition $1/M^2\to 0$ and the circular-motion condition 
$\alpha\to\theta$ yield 
\begin{equation} 
J_1-J_2\sim \sin^2 \theta J_1. 
\tag{A22} 
\end{equation}  
\subsection*{A2.2. Gravitational Torques} 
The phase difference between the major axis of the 
oval orbits and the potential valley result in angular-momentum transfer 
between the gas and the potential. 
The change in the angular momentum per unit time in the region $R\sim 
R+\Delta R$ and $\psi=0-2\pi$ is calculated by integrating the torque 
along the orbit as 
\begin{equation} 
\Delta J_z(R_0)\equiv\int_0^{2\pi}[\mbox{\boldmath $R$}\times(-\nabla\Phi)]_z 
R\Delta Rd\psi. \label{torque} 
\tag{A23} 
\end{equation} 
It was then computed numerically.

\end{document}